\documentclass[11pt]{article}
\usepackage{amssymb,latexsym,amsfonts,amsmath,epsfig,graphicx,keyval,pstricks,graphics}

\makeatletter\@addtoreset {equation}{section}\makeatother

\numberwithin{theorem}{subsection}

\DeclareMathOperator{\eps}{\varepsilon}
 
\DeclareMathOperator{\Order}{\mathcal{O}}

\begin{document}

\title{Surface gap solitons at a nonlinearity interface}

\date{\today}

\author{Tom\'{a}\v{s} Dohnal$^1$ and Dmitry Pelinovsky$^2$\footnote{\small On leave
from Department of Mathematics, McMaster
University, Hamilton, Ontario, Canada, L8S 4K1}\\
{\small $^1$ Seminar for Applied Mathematics, ETH Z\"{u}rich,
Switzerland} \\ {\small $^2$Institut f\"{u}r Analysis, Dynamik und
Modellierung, Universit\"{a}t Stuttgart, Germany} }
\maketitle

\begin{abstract}
We demonstrate existence of waves localized at the interface of
two nonlinear periodic media with different coefficients of the
cubic nonlinearity via the one-dimensional Gross--Pitaevsky
equation. We call these waves the surface gap solitons (SGS). In
the case of smooth symmetric periodic potentials, we study
analytically bifurcations of SGS's from standard gap solitons and
determine numerically the maximal jump of the nonlinearity
coefficient allowing for the SGS existence. We show that the
maximal jump vanishes near the thresholds of bifurcations of gap
solitons. In the case of continuous potentials with a jump in the
first derivative at the interface, we develop a homotopy method of
continuation of SGS families from the solution obtained via gluing
of parts of the standard gap solitons and study existence of SGS's
in the photonic band gaps. We explain the termination of the SGS
families in the interior points of the band gaps from the
bifurcation of linear bound states in the continuous non-smooth
potentials.
\end{abstract}

\section{Introduction}\label{S:intro}

We are concerned with localized waves at the interface of two
periodic nonlinear media called surface gap solitons (SGS). One of
the first publications on optical solitons propagating along
material interfaces is \cite{T80}, where the interface of a linear
and a focusing Kerr nonlinear medium is studied. In the last two
years relevant publications in the context of nonlinear optics
have dealt, for instance, with discrete surface solitons in
nonlinear waveguide arrays \cite{HSCS05,MSCSH05,SMCSHMYSS06},
surface gap solitons at the interface of a uniform and a periodic
medium with the defocusing cubic nonlinearity \cite{KVT06} and
surface vortex solitons at the interface of two periodic media
with different mean values of the refractive index and with the
saturable nonlinearity \cite{KEVT06,KT06}. One of the typical
models employed in the theory of gap solitons is the
one-dimensional nonlinear Schr\"{o}dinger equation (NLS) with
cubic nonlinearity and periodic potential called the
Gross-Pitaevsky equation.

We investigate here the existence of surface waves at the
interface of two media with identical periodic linear parts of the
refractive index and with different cubic nonlinearities. It is
known that for most photonic materials a variation in the
nonlinear part of the refractive index $n_2$ is necessarily
accompanied by a larger change in the linear part $n_0$.
Nevertheless, certain materials exhibit large variations in $n_2$
accompanied by small variations in $n_0$, see \cite{Blomer,FSW07}.
Localized states have been studied theoretically in media with
constant $n_0$ and spatially periodic $n_2$ in \cite{FSW06}.

Each of the two periodic nonlinear media supports at least two
families of standard gap solitons in every bounded non-empty
frequency gap. One family is always unstable, while the other can
be stable depending on the locations of spectral bands and
bifurcations of eigenvalues from the band edges \cite{PSK04}. The
potentially stable family looks like a single-humped envelope
soliton with exponential decay and oscillations near the central
peak. Multi-humped envelope solitons may also exist in such
periodic nonlinear media but we shall focus herein on existence of
a single-humped solution localized near the interface between the
two periodic nonlinear media.

The paper is organized as follows. Section \ref{S:background}
reviews Floquet theory for the governing Gross--Pitaevsky equation
and summarizes the results on existence of gap solitons. In
Section \ref{S:surface-small} we study the existence of SGS's for
a smooth symmetric periodic potential function and find the
maximal allowed jump in the nonlinearity coefficient between the
two media for existence of SGS's. Section \ref{S:surface-jump}
discusses bifurcations and existence of SGS's for a continuous
potential function with a derivative jump at the nonlinearity
interface. Section \ref{S:surface-conclusion} concludes the paper
with conjectures on stability of SGS's.

\section{Background: Floquet theory and gap soliton existence}
\label{S:background}

We consider the one-dimensional periodic cubic Schr\"{o}dinger
equation in the form
\begin{equation}
\label{E:per_schrod} iu_t = -u_{xx} + V(x)u - \Gamma(x) |u|^2 u,
\quad x\in\mathbb{R}, \quad t\geq 0,
\end{equation}
where $x$ and $t$ are the spatial and temporal variable
respectively, $V(x)$ is a real, continuous and $d$-periodic
potential, and $\Gamma(x) = \Gamma_{\pm}$ for $\pm x > 0$ is a
real nonlinearity coefficient with constants $\Gamma_+$ and
$\Gamma_-$. The positive values of $\Gamma(x)$ corresponds to the
focusing nonlinearity and the negative values of $\Gamma(x)$ to
defocusing nonlinearity.

We are interested in the existence of stationary solutions of
\eqref{E:per_schrod} localized near the interface at $x=0$ and
having the form
\begin{equation}\label{E:sgs_form}
u(x,t) = e^{-i\omega t} \phi(x) \quad \text{s.t.} \quad  \phi:
\mathbb{R}\rightarrow \mathbb{R}, \quad \phi \to 0 \;\;\mbox{as}
\;\; |x| \to \infty.
\end{equation}
The function $\phi(x)$ has to satisfy the second-order
non-autonomous ODE
\begin{equation}
\label{ODE} -\phi'' - \omega \phi + V(x) \phi - \Gamma(x) \phi^3 = 0,
\end{equation}
which can be cast in the Hamiltonian form with the Hamiltonian
function
\begin{equation}
\label{Hamiltonian} H[\phi] = \frac{1}{2} \left[ (\phi')^2 + \omega
\phi^2 - V(x) \phi^2 \right] + \frac{1}{4} \Gamma(x) \phi^4.
\end{equation}
Since $\Gamma(x)$ is discontinuous at $x=0$, $\phi(x)$ is a weak
solution of the ODE (\ref{ODE}) in $\phi \in C^2(\mathbb{R}_+ \cup
\mathbb{R}_-)$, such that the second derivative $\phi''(x)$ may
have a jump at $x = 0$. The continuously differentiable solution
$\phi \in C^1(\mathbb{R})$ is a critical point of the energy
functional
$$
E_{\omega}[\phi] = \frac{1}{2} \int_{\mathbb{R}} \left[ -|\phi'|^2
+ \omega |\phi|^2 - V(x) |\phi|^2 \right] dx + \frac{1}{4}
\int_{\mathbb{R}} \Gamma(x) |\phi|^4  dx,
$$
such that the first variation of $E_{\omega}'[\phi]$ recovers the
ODE (\ref{ODE}).

Replacing $t$ by $z$ in \eqref{E:per_schrod} and \eqref{E:sgs_form},
the $x-$localized solution $u(x,z)$ can be viewed as a spatial soliton propagating
along the direction $z$ and localized in the transverse direction
$x$ (localization in the third spatial direction $y$ is assumed to
be achieved via total internal reflection). The parameter $\omega$
plays a role of the propagation constant. As we show below, the
localized solutions of the ODE \eqref{ODE} decay exponentially as
$|x| \to \infty$ only if $\omega$ belongs to the frequency gaps in
the continuous spectra of the operator $L := -\partial_{xx} + V(x)$
called the photonic band gaps. To do so, we recall the basic Floquet
theory (see \cite{Eastham73,Magnus_Win_66}) for the Hill's equation
\begin{equation}\label{E:spectral_prob}
L \psi(x) =  -\psi''(x) + V(x) \psi(x) = \omega \psi(x), \qquad x
\in \mathbb{R}.
\end{equation}
The bounded solutions $\psi(x)$ of the Hill's equation
\eqref{E:spectral_prob} are usually called Bloch functions. Given
a real, continuous and $d$-periodic potential $V(x)$, bounded
solutions $\psi(x)$ exist for $\omega$ in a union of (possibly
disjoint) spectral bands from
\[
\Sigma := [\omega_0,\omega_1] \cup [\omega_2,\omega_3] \cup  [\omega_4,\omega_5] \cup \ldots,
\]
where $\omega_{2n-2} < \omega_{2n-1} \leq \omega_{2n}$, $n \in
\mathbb{N}$ and $\omega_n \rightarrow \infty$ as $n \rightarrow
\infty$.  The set $\Sigma$ represents the complete (purely
continuous) spectrum of the operator $L$ \cite{Eastham73}. We
shall {\em assume} for simplicity that all spectral bands are
disjoint with $\omega_{2n-1} < \omega_{2n}$, $n \in \mathbb{N}$,
such that all finite frequency gaps are non-empty.

For a fixed $\omega$ in the interior point of $\Sigma$, both
fundamental solutions of the second-order ODE
\eqref{E:spectral_prob} are quasi-periodic in $x$ and have the
representation $\psi = p_{\pm}(x) e^{\pm i k x}$, where
$p_{\pm}(x) = p_{\pm}(x+d)$ and $k \in
\left[0,\frac{\pi}{d}\right]$. The parameter $k$ parameterizes the
frequency parameter $\omega$, such that we shall use notation
$\omega = \omega_{2n,2n+1}(k)$ for the spectral band in $\omega
\in [\omega_{2n},\omega_{2n+1}]$. If the $n$-th band is separated
from the $(n+1)$-th band (i.e. $\omega_{2n-1} < \omega_{2n}$ and
$\omega_{2n+1} < \omega_{2n+2}$), then $\omega_{2n,2n+1}'(k) = 0$
and $\omega_{2n,2n+1}''(k) \neq 0$ at the end points $k = 0$ and
$k = \frac{\pi}{d}$ \cite{Kohn}.

When $\omega = \omega_n$, one of the solutions $\psi = \psi_n(x)$
is either $d$-periodic (corresponding to $k = 0$) or
$d$-antiperiodic (corresponding to $k = \frac{\pi}{d}$) and the
other fundamental solution $\psi(x)$ grows linearly in $x$. For a
fixed $\omega \in \mathbb{R} \setminus \Sigma$ the two fundamental
solutions of \eqref{E:spectral_prob} grow exponentially either in
$x$ or $-x$ and have the representation $\psi = u_{\pm}(x) e^{\pm
\kappa x}$, where $u_{\pm}(x)$ is either periodic or anti-periodic
and $\kappa = \kappa(\omega) \in \mathbb{R}_+$. The functions
$u_{\pm}(x)$ are periodic (anti-periodic if the bounded solutions
$\psi_n(x)$ are periodic (anti-periodic) at the band edges
$\omega_{2n-1}$ and $\omega_{2n}$, which surround the band gap.

Suppose that $\phi(x)$ is a localized solution of the ODE
(\ref{ODE}). It is then obvious from the linearized analysis that
the solution $\phi(x)$ decays exponentially as $|x| \to \infty$
only if $\omega \in \mathbb{R} \setminus \Sigma$. It was shown
under fairly general assumptions (see \cite{PSK04} and references
therein) that the families of gap solitons of the ODE (\ref{ODE})
with constant coefficient $\Gamma(x) = \Gamma_0$ undertake a local
bifurcation from all points $\omega = \omega_{2m}$, $m \geq 0$ to
the left if $\Gamma_0 > 0$ and from all points $\omega =
\omega_{2m+1}$, $m \geq 0$ to the right if $\Gamma_0 < 0$ (the
term {\em local bifurcation} means that $\| \phi \|_{L^{\infty}}
\to 0$ as $\omega \to \omega_n$). This conjecture was rigorously
proved in \cite{pankov}, where existence of exponentially decaying
gap solitons in $H^1(\mathbb{R})$ was confirmed in every finite
frequency gap $\omega \in (\omega_{2m-1}, \omega_{2m})$, $m \in
\mathbb{N}$ and in the semi-infinite frequency gap $\omega <
\omega_0$ for $\Gamma_0 > 0$. We use this result but simplify our
consideration by working with the class of symmetric potentials $V
= V_0(x)$, where $V_0(-x) = V_0(x)$ on $x \in \mathbb{R}$. In
particular, we shall perform numerical computations with
\begin{equation}
\label{potential-example} V_0(x) = \sin^2 \left( \frac{\pi x}{d}
\right), \quad d=10,
\end{equation}
which has a minimum at $x = 0$, i.e. at our interface location.
The spectral bands and gaps of $V_0(x)$ are approximated
numerically from the Hill's equation (\ref{E:spectral_prob}). For
instance, the first five band edges of the potential
(\ref{potential-example}) are located as follows
$$
\omega_0 \approx 0.283, \quad \omega_1 \approx 0.291, \quad
\omega_2 \approx 0.747, \quad \omega_3 \approx 0.843, \quad
\omega_4 \approx 1.057.
$$
As partly seen in Fig. 1 of \cite{PSK04}, for $V_0(x) = \sin^2(\pi x/10)$ the Bloch functions
$\psi = \psi_n(x)$ at the first eight band edges $\omega = \omega_n$, $n \in \{0,1, \ldots, 7\}$ have the following symmetry properties:
\begin{equation}\label{symmetry-eigenfunctions}
\begin{split}
&\psi_n(-x) = \psi_n(x), \; n \in
\{ 0,1,4,5 \},\\
&\psi_n(-x) = - \psi_n(x),  \; n \in
\{ 2,3,6,7 \}.
\end{split}
\end{equation}
Clearly, all $\psi_n(x)$ must be even or odd since the Hill's equation (\ref{E:spectral_prob}) is symmetric with respect
to reflection $x \mapsto -x$ and admits only one linearly
independent bounded eigenfunction $\psi = \psi_n(x)$ at $\omega =
\omega_n$. Nevertheless, for other even potentials $V_0(x)$ the ordering between even and odd Bloch functions can be different than in \eqref{symmetry-eigenfunctions} and the only statement about this ordering that is valid for general even $V_0(x)$ can be deduced from Theorem 3.1.1 in \cite{Eastham73}, which says that two subsequent even eigenfunctions have to be followed by an odd one and vice versa. We refrain from such a discussion here concentrating on the potential \eqref{potential-example}, for which symmetries \eqref{symmetry-eigenfunctions} hold.

In contrast to the ordering between even and odd Bloch functions, the ordering between $d-$periodic and $d-$antiperiodic Bloch functions is unique for all $d-$periodic $V_0(x)$ (not necessarily even). By the trace of the monodromy matrix \cite{Eastham73}, the
periodic functions $\psi_n(x)$ correspond to the set $n \in S_+$
with $S_+ = \{ 0, 3, 4, 7, 8,...\} = \{0, 4k-1, 4k; k\in \mathbb{N} \}$ and the anti-periodic functions
$\psi_n(x)$ correspond to the set $n \in S_-$ with $S_- = \{ 1,2,5,6,...\} = \{4k-3, 4k-2; k\in \mathbb{N} \}$. 

By Sturm's Theorem (see Theorem 3.1.2 in \cite{Eastham73}) for general $d-$periodic $V_0(x)$
the periodic functions $\psi_n(x)$ with $n \in S_+$ have no 
nodes on $x \in [0,d)$ for $n = 0$, 
two nodes for $n = 3,4$, four nodes for $n = 7,8$, etc. 
Similarly, the anti-periodic functions $\psi_n(x)$ with $n \in S_-$ have 
one node on $x \in [0,d)$ for $n = 1,2$, 
three nodes for $n = 5,6$, etc.

Altogether, this set of facts about Bloch functions at the first eight band edges for the potential \eqref{potential-example} is summarized in Table \ref{T:Bloch_prop}.

\begin{table}[h!]
\begin{center}
\begin{tabular}{||c||c|c|c|c|c|c|c|c||}
\hline
$n$ & 0 & 1 & 2 & 3 & 4 & 5 & 6 & 7\\
\hline
symmetry & even & even & odd & odd & even & even & odd & odd\\
\hline
periodicity &  $S_+$ & $S_-$&$S_-$ & $S_+$ & $S_+$ &$S_-$ & $S_-$ &$S_+$ \\
\hline
\# nodes on $[0,d)$ & 0 & 1 & 1 & 2 & 2 & 3 & 3 & 4\\\hline
sign of $\Gamma_0$ for local bifurcation & 1 & -1 & 1 & -1 & 1 & -1 & 1 & -1\\
\hline
\end{tabular}
\vspace{0.25cm} \caption{Properties of the Bloch functions
$\psi_n(x)$ and gap soliton bifurcations at the first eight band
edges of the even potential $V_0(x)=\sin^2(\pi x/10)$.}\label{T:Bloch_prop}
\end{center}
\end{table}

Let $\phi_0(x)$ be a single-humped solution of the ODE (\ref{ODE})
with $\Gamma(x) = \Gamma_0$ and $V(x) = V_0(x)=\sin^2(\pi x/10)$ which bifurcates
from the band edge $\omega = \omega_n$. By the local bifurcation
theory \cite{PSK04}, it inherits the symmetry properties
(\ref{symmetry-eigenfunctions}) of the Bloch function $\psi_n(x)$.
Therefore, $\phi_0(-x) = \phi_0(x)$ for branches of gap solitons
to the left of $\omega_n$ with $n = \{0,4\}$ (for $\Gamma_0
> 0$) and to the right of $\omega_n$ with $n = \{1,5\}$ (for $\Gamma_0 < 0$), while
$\phi_0(-x) = -\phi_0(x)$ for branches of gap solitons to the left
of $\omega_n$ with $n = \{2,6\}$ (for $\Gamma_0 > 0$) and to
the right of $\omega_n$ with $n = \{ 3,7\}$ (for $\Gamma_0 <
0$). See Figs. 2-3 in \cite{PSK04} for gap solitons $\phi_0(x)$ in
the potential (\ref{potential-example}).

In this paper, we shall consider existence of surface gap solitons
in the ODE (\ref{ODE}) with piecewise constant coefficient
$\Gamma(x) = \Gamma_{\pm}$ for $\pm x > 0$ and potential $V(x)$ of
the following two classes:
\begin{equation}
\label{potentials} \mbox{(i)} \; V = V_0(x), \qquad \mbox{(ii)} \;
V = V_0(x - \delta) \chi_{(-\infty,0)} + V_0(x +
\delta)\chi_{[0,\infty)},
\end{equation}
where $\chi_{[a,b]} = 1$ on $x \in [a,b]$ and zero otherwise,
while $0 < \delta < d$. Here $V_0(x)$ is a smooth, even,
$d$-periodic function on $x \in \mathbb{R}$. We note that $V(x)$
in (ii) is continuous and even on $x \in \mathbb{R}$ but smooth
and periodic only on each $\pm x>0$.

One can develop a general shooting method for numerical
approximations of SGS's from the condition that a localized
solution $\phi(x)$ of the second-order ODE (\ref{ODE}) with
$\omega \in (\omega_{2m-1},\omega_{2m})$, $m \in \mathbb{Z}$
decays to zero at infinity according to two fundamental solutions
$p_{\pm}(x) e^{\mp \kappa x}$ as $x \to \pm \infty$, where $\kappa
= \kappa(\omega)$ is a positive number. Solving the ODE
(\ref{ODE}) with $\Gamma(x) = \Gamma_+$ for a general initial
value $\phi(0)$ and $\phi'(0)$ to $x > 0$ and the same ODE with
$\Gamma(x) = \Gamma_-$ to $x < 0$, one can construct a
continuously differentiable solution $\phi(x)$ on $x \in
\mathbb{R}$ which decays to zero as $x \to \pm \infty$ if and only
if the projections to the growing fundamental solutions
$p_{\pm}(x) e^{\pm \kappa x}$ are zero at infinity. The system of
two constraints for two initial values constitute a well-posed
problem of numerical analysis. However, practical implementations
of this algorithm are unclear as the shooting method may depend
sensitively on starting approximations of the initial value and
may require long computations time to search through all
appropriate initial values. In addition, the ODE solvers of the
shooting method may develop numerical instabilities in
approximations of growing solutions.

Due to these reasons, we shall develop an alternative view on
numerical approximations of SGS's, starting with local bifurcation
analysis and using the homotopy continuation method to trace the
solution families along parameters $\omega$, $\Gamma_{\pm}$ and
$\delta$. Using these analytical and numerical results, we have
obtained the following main results of the article.

\begin{enumerate}
\item We prove analytically that any gap soliton for $\Gamma_+ =
\Gamma_-$ can be continued to the SGS for
sufficiently small $|\Gamma_+ - \Gamma_-|$ under a non-degeneracy
assumption.

\item We prove analytically that the maximal difference $|\Gamma_+
- \Gamma_-|$ leading to SGS existence converges to $0$ when
$\omega$ approaches the band edge which features the local
bifurcation of a gap soliton.

\item Surface gap solitons are computed numerically when the
potential $V(x)$ is given by (\ref{potentials})(i), and the
maximal $|\Gamma_+ - \Gamma_-|$  allowing their existence is
found. Our numerical results confirm the analytical results (1-2)
above.

\item Existence of SGS's for $V(x)$ in (\ref{potentials})(ii) with
$\Gamma_+ > 0$ and $\Gamma_- < 0$ is studied. We show numerically
that local bifurcations may occur from a countable set of points
in the parameter domain $(\omega,\delta) \in
(\omega_{2m-1},\omega_{2m})\times(0,d)$, $m \in \mathbb{N}$.

\item We compute numerically the points of local bifurcation of
SGS's for the potential (\ref{potentials})(ii) and use the
homotopy continuation of the bifurcating solution. As a result, we
show that the family of SGS's exists typically in a subset of the
plane $(\omega,\delta)$.

\item We show analytically that the termination of families of
SGS's for the potential (\ref{potentials})(ii) is related to
existence of linear bound states for the non-smooth potential.
\end{enumerate}

Results (1-3) are reported in Section \ref{S:surface-small} and
results (4-6) are described in Section \ref{S:surface-jump}.

\section{Bifurcations of surface gap solitons for smooth potentials}
\label{S:surface-small}

In this section we study continuation of SGS's from gap solitons
existing for $\Gamma_+ = \Gamma_-$ in the case of a smooth
potential function $V(x)$. A prototypical example of such
potential is the symmetric function $V_0(x)$ in
(\ref{potentials})(i).

\subsection{Existence of Bifurcations from Gap Solitons}

Let $\gamma = (\Gamma_+ + \Gamma_-)/2$ and $\nu = (\Gamma_+ -
\Gamma_-)/2$. Then, the ODE (\ref{ODE}) can be rewritten in the
form
\begin{equation}
\label{small-ODE} F(\phi,\nu) = -\phi'' - \omega \phi + V(x) \phi
- \gamma \phi^3 - \nu \; {\rm sign}(x) \phi^3 = 0,
\end{equation}
where $F(\phi,\nu) : H^1(\mathbb{R}) \times \mathbb{R} \mapsto
H^{-1}(\mathbb{R})$ is a nonlinear operator acting on a function
$\phi(x)$ in space $\phi \in H^1(\mathbb{R})$ and parameter $\nu
\in \mathbb{R}$.

We assume that there exists a solution $\phi_0(x) \in
H^1(\mathbb{R})$ for $\omega \in \mathbb{R} \setminus \Sigma$ and
some $\gamma$ and $V(x)$, such that $F(\phi_0,0) = 0$. The
Jacobian $D_{\phi} F(\phi_0,0)$ is given by the Schr\"{o}dinger
operator $\mathcal{L}: H^2(\mathbb{R}) \mapsto L^2(\mathbb{R})$,
where
\begin{equation}
\mathcal{L} = - \partial_x^2 - \omega + V(x) - 3 \gamma
\phi_0^2(x).
\end{equation}
Since $\omega \in \mathbb{R} \setminus \Sigma$, we have
$\phi_0^2(x) \to 0$ exponentially fast as $|x| \to \infty$, such
that the term $-3 \gamma \phi_0^2(x)$ is a relatively compact
perturbation to the unbounded operator $L-\omega$, where $L =
-\partial_x^2 + V(x)$. By a standard argument (see Corollary 2 in
Section XIII.4 in \cite{Kato}), the essential spectrum of
$\mathcal{L}$ and $(L - \omega)$ coincide. Since $\omega \in
\mathbb{R} \setminus \Sigma$, the zero point is isolated from the
essential spectrum of $\mathcal{L}$. If we further assume that
$\mathcal{L}$ has the trivial kernel in $H^1(\mathbb{R})$, then
$\mathcal{L}$ is invertible on $L^2(\mathbb{R})$. Since the
translational invariance is broken if $V(x)\neq 0$, $\mathcal{L}$
generally has the trivial kernel, unless a bifurcation of branches
of gap solitons occur. By the standard analysis based on the
Implicit Function Theorem, there exists a unique smooth
continuation of $\phi_{\nu}(x)$ from $\phi_0(x)$ in
$H^1(\mathbb{R})$ for sufficiently small $\nu$, such that
$F(\phi_{\nu},\nu) = 0$ and $\phi_{\nu}(x) \to \phi_0(x)$ in
$H^1(\mathbb{R})$ as $\nu \to 0$.

In other words, we have proved above that if a gap soliton exists
for $\Gamma_+ = \Gamma_-$ and $\omega \in \mathbb{R} \setminus
\Sigma$ and the linearized operator $\mathcal{L}$ is
non-degenerate, then the gap soliton is uniquely continued into
the SGS for small non-zero $|\Gamma_+ - \Gamma_-|$. We confirm
this prediction via numerical analysis of the ODE (\ref{ODE}) with
$V(x)$ in (\ref{potentials})(i) for $\omega$ taken in the
semi-infinite band gap and the first two finite gaps. Numerical
approximations of $\phi_0(x)$ for $\Gamma_+ = \Gamma_-$ are
obtained from the Newton--Raphson iterations and the homotopy
continuation method. The initial guess for the Newton's iteration
is taken from an asymptotic expansion leading to the NLS
approximation \cite{PSK04} when $\omega$ is close to the local
bifurcation threshold $\omega_n$. After a successful convergence
for one such $\omega$ we use a standard homotopy continuation and
generate a family of gap solitons $\phi_0(x)$ parameterized by
$\omega$. The discretization of the ODE \eqref{ODE} is based on a
fourth order central difference approximation of $\partial_{xx}$
on a truncated domain with zero Dirichlet boundary conditions.

\subsection{Numerical Computations of Surface Gap Solitons}

We now proceed to construct SGS's, i.e. solutions
$\phi(x)$ of the second-order ODE (\ref{ODE}) with $\Gamma_+ \neq
\Gamma_-$. When $\phi_0(x)$ is obtained for a given value of
$\omega$, we can apply the numerical homotopy continuation of the
solution by deviating $\Gamma_-$ from $\Gamma_+$. At each step,
the SGS $\phi(x)$ is thus found via Newton's
iterations. The final value of $\Gamma_-$, up to which the
iteration converges, is denoted by $\Gamma_*$.

\begin{figure}
\epsfig{figure = 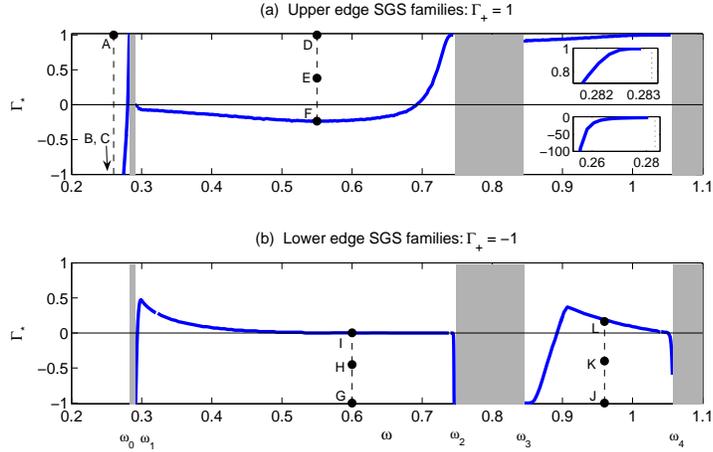,scale=.42}
\caption{The values of $\Gamma_*$ for SGS's originating from
symmetric GS families of the first three frequency gaps of
$V(x)=\sin^2(\pi x/10)$. In (a) the upper inset zooms in and the
lower inset zooms out on the graph in the semiinfinite gap. The
points $A-L$ are referenced in Fig. \ref{F:SGS_profiles}.}
\label{F:Gam_last}
\end{figure}

Fig. \ref{F:Gam_last} shows the values of $\Gamma_*$ for $\Gamma_+
= +1$ (a) and $\Gamma_+ = -1$ (b). The computational tolerance in
$\Gamma_*$ is $0.006$ inside the band gaps and $0.002$ near the
band edges. In the case $\Gamma_+ = 1$, local bifurcations of
small-amplitude gap solitons occur from the lower band edges. Fig.
\ref{F:Gam_last}(a) shows that the SGS's exist in the
semi-infinite gap, as well as in the first two frequency gaps. In
the case $\Gamma_+ = -1$, local bifurcations of gap solitons occur
from the upper band edges. Fig. \ref{F:Gam_last}(b) shows that the
SGS's exist in the first and second frequency gaps. The two insets
of Fig. \ref{F:Gam_last}(a) show that $\Gamma_*$ decreases fast as
$\omega$ moves away from the edge of the first band and that the
convergence $\Gamma_* \uparrow 1$ as $\omega \uparrow \omega_0$ is
smooth. We further see from Fig. \ref{F:Gam_last} that the
interval of existence shrinks as $\omega$ approaches the value
$\omega_n$ for any band edge, where gap solitons undertake a local
bifurcation. In addition, the interval of existence is extremely
large in the semi-infinite gap $(-\infty,\omega_0)$, but it
becomes narrow in the finite gaps $(\omega_{2m-1},\omega_{2m})$
for $m \geq 1$.

For comparison, the family of SGS's in the gap
$(\omega_1,\omega_2)$ exists for $-0.24 < \Gamma_* < 1$ in the
case $\Gamma_+ = +1$ and $-1 < \Gamma_* < 0.47$ in the case
$\Gamma_+ = -1$. The family of SGS's in the gap
$(\omega_3,\omega_4)$ exists in a very narrow region of $0.92 <
\Gamma_* < 1$ in the case $\Gamma_+ = +1$ and in a bigger interval
$-1 < \Gamma_* < 0.37$ in the case $\Gamma_+ = -1$ (similarly to
that in the first gap).

\begin{figure}
\epsfig{figure = 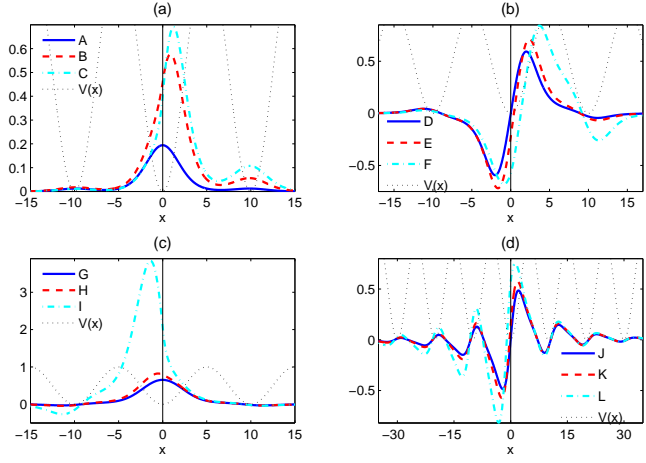,scale=.42}
\caption{The profiles of SGS's corresponding to the points $A-L$ in
Fig. \ref{F:Gam_last}. Values of $\omega$ are $A-C$: 0.26, $D-F$:
0.55, $G-I$: 0.6, $J-L$: 0.96. Values of $\Gamma_-$ are $A$: 1, $B$: -3.9,
$C$: -15.3, $D$: 1, $E$: 0.38, $F$: -0.235, $G$: -1, $H$: -0.45, $I$: 0.002, $J$:
-1, $K$: -0.4, $L$: 0.164.} \label{F:SGS_profiles}
\end{figure}

Fig. \ref{F:SGS_profiles} shows profiles of SGS's which correspond
to the twelve points labeled $A-L$ in Fig. \ref{F:Gam_last}. The
full lines correspond to the gap solitons from which the homotopy
in $\Gamma_-$ is started (i.e. points $A, D, G$ and $J$). Clearly, the
total power and maximum amplitude of the SGS increase as
$|\Gamma_+-\Gamma_-|$ increases. Also notice that the profiles
become more concentrated on the half $x>0$ in the case $\Gamma_+ =
+1$ [see Fig. \ref{F:SGS_profiles} (a-b)] and on the half $x < 0$
in the case $\Gamma_+ = -1$ [see Fig. \ref{F:SGS_profiles} (c-d)]
as $|\Gamma_+-\Gamma_-|$ increases. This is in accord with the law
of refraction: when $\Gamma_+ = +1$ and $\Gamma_-$ decreases from
$1$, the half $x>0$ becomes relatively more focusing and therefore
attracts more energy of the soliton, while when $\Gamma_+ = -1$
and $\Gamma_-$ increases from $-1$, the situation is opposite.

\subsection{Asymptotic Analysis near Gap Soliton Bifurcation Points}

We shall explain now why the existence interval shrinks to zero
when $\omega$ approaches the value $\omega_n$ where a local
bifurcation of gap solitons occurs. As $\omega \to \omega_n$, we
have  $\| \phi_0 \|_{L^{\infty}} \to 0$ and $\mathcal{L} \to (L -
\omega_n)$. Since the operator $(L-\omega_n)$ is not invertible,
the Implicit Function Theorem can not be used and the solution
$\phi_0(x)$ can not be continued beyond $\nu = 0$. In order to
give a more precise explanation of this phenomenon, we adopt the
NLS approximation for local bifurcation of gap solitons from
\cite{PSK04} (see also review in \cite{Pelin}). In particular, we
consider an asymptotic solution to the ODE (\ref{ODE}):
\begin{equation}\label{E:expansion}
\begin{split}
& \omega = \omega_n + \eps^2 \Omega + \Order(\eps^4),\\
& \phi(x) = \eps A(X) \psi_n(x)+ \eps^2 A'(X) \tilde{\psi}_n(x) +
\eps^3 \phi^{(3)}(x,X) + \Order(\eps^4),
\end{split}
\end{equation}
where $X=\eps x$, $\eps <<1$, the function $A(X)$ and parameter
$\Omega$ are defined below; $\psi_n$ and $\tilde{\psi}_n$ are
the $d$-periodic (or $d$-antiperiodic) Bloch function and generalized Bloch functions
respectively of the Hill's
equation (\ref{E:spectral_prob}) for $\omega = \omega_n$, such
that
\begin{equation}\label{E:Bloch_fn_eqs}
(L-\omega_n) \psi_n = 0, \qquad (L-\omega_n)\tilde{\psi}_n = 2
\psi_n'.
\end{equation}
The correction term $\phi^{(3)}(x,X)$ at $\Order(\eps^3)$ solves the
non-homogeneous problem
\begin{equation}
\label{inhom-equat-3} (L - \omega_n) \phi^{(3)} = \Omega A \psi_n + A''
\psi_n + 2 A'' \tilde{\psi}'_n + \Gamma(X) A^3 \psi_n^3.
\end{equation}
To ensure boundedness of $\phi^{(3)}(x,X)$ with respect to the variable
$x$, and, hence, legitimacy of the expansion \eqref{E:expansion},
one has to apply the Fredholm alternative which imposes the
orthogonality condition of the right-hand-side of
(\ref{inhom-equat-3}) with respect to $\psi_n(x)$ on $x \in [0,d]$.
The orthogonality condition is written as follows
\begin{equation}\label{E:An_equ}
\Omega A + \mu A'' + \rho \Gamma(X) A^3 = 0,
\end{equation}
where
$$
\mu = 1 + 2 \frac{(\tilde{\psi}'_n, \psi_n)}{(\psi_n,\psi_n)}, \qquad
\rho = \frac{(\psi_n^2,\psi_n^2)}{(\psi_n,\psi_n)},
$$
and we have used the standard $L^2$ inner product $(\cdot,\cdot)$
over one period $x\in[0,d]$. It is shown in \cite{PSK04} that $\mu
= \frac{1}{2} \omega''_{2n,2n+1}(k)$ with either $k = 0$ or $k =
\frac{\pi}{d}$ at the point $\omega = \omega_n$, where
$\omega_{2n,2n+1}(k)$ is the dispersion relation between $\omega
\in [\omega_{2n},\omega_{2n+1}]$ and $k \in [0,\frac{\pi}{d}]$.

Due to the nature of the nonlinearity interface, the function
$\Gamma(X)$ is the same as $\Gamma(x)$, i.e. $\Gamma(X) =
\Gamma_{\pm}$ for $\pm X > 0$. We shall prove that no localized
solution of the ODE \eqref{E:An_equ} exists under the condition
$\Gamma_- \neq \Gamma_+$. Indeed, consider the Hamiltonian of the
ODE (\ref{E:An_equ}):
\begin{equation}
\label{Hamiltonian-A-equation} H[A] = \frac{1}{2} \left[ \mu (A')^2
+ \Omega A^2 \right] + \frac{1}{4} \rho \Gamma(X) A^4.
\end{equation}
If $A(X)$ solves the ODE (\ref{E:An_equ}), then
$$
\frac{d}{dX} H[A(X)] = \frac{1}{4} \rho \Gamma'(X) A^4(X) =
\frac{1}{4} \rho (\Gamma_+ - \Gamma_-) \delta(X) A^4(X),
$$
where $\delta(X)$ is the Dirac delta-function. If $A(X)$ is a
localized solution on $X \in \mathbb{R}$, then the integration on
$X \in \mathbb{R}$ gives the constraint
$$
0=\lim_{x \to +\infty} H[A(X)] - \lim_{x \to -\infty} H[A(X)] =
\frac{1}{4} \rho (\Gamma_+ - \Gamma_-) A^4(0),
$$
since $H[A(X)] \to 0$ if $A(X),A'(X) \to 0$ as $|X| \to \infty$.
Therefore, $A(0) = 0$ if $\Gamma_+ \neq \Gamma_-$. Consider now
$H[A(X)]$ on $X > 0$. It is clear from the decaying conditions as
$X \to \infty$ that $H[A(X)] = {\rm const} = 0$, which together
with the fact $A(0)=0$ leads to $0 = \lim\limits_{X \downarrow 0}
H[A(X)] = \frac{1}{2} \mu |A'(0)|^2$, such that $A'(0) = 0$. The
only solution of the ODE (\ref{E:An_equ}) with $A(0) = A'(0) = 0$
is the zero solution $A(X) \equiv 0$.

If $\Gamma_+ = \Gamma_- = \Gamma_0$ and ${\rm sign}(\mu) = {\rm
sign}(\rho \Gamma_0) = -{\rm sign}(\Omega)$, the ODE
(\ref{E:An_equ}) has the standard sech-soliton decaying as $|X|
\to \infty$. However, the result above shows that the sech-soliton
with $\Gamma_+ = \Gamma_-$ cannot be homotopically continued to a
decaying solution of (\ref{E:An_equ}) for $\Gamma_+ \neq \Gamma_-$.
This proves that $\Gamma_* \rightarrow \Gamma_+$ as $\omega\rightarrow \omega_n$
where $\omega_n$ is a local bifurcation value.

\section{Bifurcations of surface gap solitons for nonsmooth potentials}
\label{S:surface-jump}

In this section, we study local bifurcations of solutions of the
ODE (\ref{ODE}) when $V(x)$ is a continuous function with the jump
in the first derivative at the nonlinearity interface. The
prototypical example of such potentials is given by
(\ref{potentials})(ii), where $V_0(x)$ is an even potential (in
our numerical computations we use $V_0$ from \eqref{potential-example}). We
shall consider the existence of SGS's under the normalization
$\Gamma_+ = - \Gamma_- = +1$.

\subsection{Surface Gap Soliton Numerical Construction via Gluing}\label{S:gluing}

The point $(\delta_*,\omega_*)$ in the parameter domain $\delta
\in (0,d)$ and $\omega \in (\omega_{2m-1},\omega_{2m})$, $m \in
\mathbb{N}$ is defined to be a point of a local bifurcation of
SGS's according to the following 2-step algorithm. \\

\noindent
\textit{(i) Construction of Continuous Solutions}

Let $\phi_\pm(x;\omega)$ denote the family of single--humped gap
solitons parameterized by $\omega \in (\omega_{2m-1},\omega_{2m})$
and centered at $x=0$ corresponding to the equation \eqref{ODE}
with $\Gamma(x)\equiv \Gamma_\pm$ respectively. These families
bifurcate from the points $\omega = \omega_{2m}$ for $\Gamma_+
> 0$ and $\omega = \omega_{2m-1}$ for $\Gamma_- < 0$. In order to
find continuous solutions, we now study for each fixed $\delta
\in (0,d)$ the two functions
$$
f_A(\omega) = \phi_-(-\delta;\omega) - \phi_+(\delta;\omega), \quad
f_B(\omega) = \phi_-(-\delta;\omega) + \phi_+(\delta;\omega)
$$
and find their zeros denoted by
$\omega_{A,B}=\omega_{A,B}(\delta)$, respectively. For each
$\delta$ existence of zeros of either $f_A(\omega)$ or
$f_B(\omega)$ is guaranteed by continuity of $\phi_\pm$ as
functions of $\omega$ and by the fact that
$\phi_-(-\delta;\omega_{2m-1}) = \phi_+(\delta;\omega_{2m})=0$ and
$\phi_-(-\delta;\omega_{2m}) \neq 0$,
$\phi_+(\delta;\omega_{2m-1}) \neq 0$. Moreover, several zeros of
these functions may occur for the same $\delta$.

When a zero $\omega_A(\delta)$ or  $\omega_B(\delta)$ is found, a
$\delta$-parameterized family of continuous solutions
$\phi_A(x;\delta)$ or $\phi_B(x;\delta)$, respectively, is constructed
by gluing two individual gap
solitons
\begin{equation} \label{function-soliton}
\begin{split}
&\phi_A(x;\delta) =
\phi_-(x-\delta;\omega_A) \chi_{(-\infty,0)} +
\phi_+(x+\delta;\omega_A) \chi_{[0,\infty)}\\
&\phi_B(x;\delta) =
\phi_-(x-\delta;\omega_B) \chi_{(-\infty,0)} -
\phi_+(x+\delta;\omega_B) \chi_{[0,\infty)}
\end{split}
\end{equation}
The functions $\phi_{A,B}(x;\delta)$ decay as $|x|\rightarrow \infty$
and are smooth in $x$ everywhere except at the nonlinearity
interface $x=0$, where they generally have a jump in the first derivative.

Note that it is important to consider both $\phi_A$ and $\phi_B$ due to the
sign invariance of the ODE (\ref{ODE}). Each sign
produces a branch of continuous solutions of the ODE (\ref{ODE}).

\begin{figure}
\epsfig{figure =
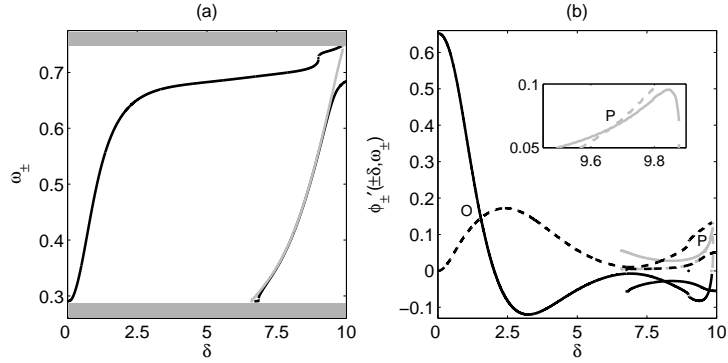,scale=.48}
\caption{Two-step search for $(\omega_*,\delta_*)$ in the gap
$(\omega_1,\omega_2)$. (a) Result of step (i) - parametrization of
the families of continuous solutions \eqref{function-soliton}:
black line $\omega_A(\delta)$, gray line $\omega_B(\delta)$; (b)
step (ii) - search for $\delta_*$: solid black
$\phi_+'(\delta;\omega_A)$, dashed black
$\phi_-'(-\delta;\omega_A)$, solid gray
$-\phi_+'(\delta;\omega_B)$ and dashed gray
$\phi_-'(-\delta;\omega_B)$. Labeled points correspond to $C^1$
SGS's.} \label{F:om_delta_gap2}
\end{figure}

\begin{figure}
\epsfig{figure =
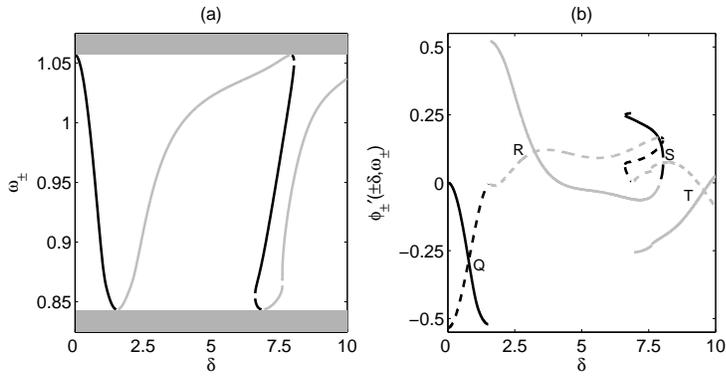,scale=.48}
\caption{Analogous to Fig. \ref{F:om_delta_gap2} but for the gap $(\omega_3,\omega_4)$.} \label{F:om_delta_gap3}
\end{figure}

Figs. \ref{F:om_delta_gap2} (a) and \ref{F:om_delta_gap3} (a) present
the numerically computed $\omega_{A,B}(\delta)$ in the gaps
$(\omega_1,\omega_2)$ and $(\omega_3,\omega_4)$ respectively. The
lack of smoothness in the curves in these figures is due to
an insufficient resolution in the search algorithm and can be
corrected with a finer resolution. Note that when
$\omega_{A,B}(\delta)$ is multiple valued as seen in Fig.
\ref{F:om_delta_gap2} (a) and \ref{F:om_delta_gap3} (a), we may
have several decaying solutions $\phi_A(x)$ and/or $\phi_B(x)$
for the same $\delta$.\\

\noindent
\textit{(ii) Construction of $C^1$ Surface Gap Solitons}

Next, we search for continuously differentiable solutions within
the above family $\phi_{A,B}(x;\delta)$. To ensure the continuity of the
first derivative of $\phi(x;\delta)$ at $x=0$, we search for zeros of the two
functions
$$
g_A(\delta) = \phi_-'(-\delta;\omega_A) -
\phi_+'(\delta;\omega_A), \quad
g_B(\delta) = \phi_-'(-\delta;\omega_B) +
\phi_+'(\delta;\omega_B).
$$
If a zero of either $g_A(\delta)$ or $g_B(\delta)$, denoted by
$\delta_*$, exists, then the function $\phi_A(x;\delta_*)$ or
$\phi_B(x;\delta_*)$, respectively, in (\ref{function-soliton})
has a continuous first derivative across the point $x = 0$.  Figs.
\ref{F:om_delta_gap2} (b) and \ref{F:om_delta_gap3} (b) present
the numerical results on computing $\delta_*$. The labelled
intersection points $O, P, Q, R,S$ and $T$ correspond to zeros of
$g_{A,B}(\delta)$. They are found as intersection points of solid
and dashed curves of the same color. The solid black line shows
the plot of $\phi_+'(\delta;\omega_A)$  and the dashed black line
shows $\phi_-'(-\delta;\omega_A)$. Similarly, the solid gray line
plots $-\phi_+'(\delta;\omega_B)$ and the dashed gray line plots
$\phi_-'(-\delta;\omega_B)$. Therefore, an intersection of a solid
black and a dashed black line (points $O,Q,S$) gives zeros
$\delta_*$ of $g_A(\delta)$ and, thus, a $C^1$ SGS
$\phi_A(x;\delta_*)$. Similarly, an intersection of a solid gray
and a dashed gray line (points $P,R,T$) gives zeros  $\delta_*$ of
$g_B(\delta)$ and, thus, a $C^1$ SGS $\phi_B(x;\delta_*)$.

Table \ref{T:bifurcations} shows the approximate computed values
of $\delta_*$ and corresponding $\omega_* =
\omega_{A,B}(\delta_*)$ at the points $O-T$ for branches $A, B$
of solutions given by \eqref{function-soliton}. Note that
additional points $(\delta_*,\omega_*)$ can be obtained by
generalizing the above functions $f_{A,B}$ and $g_{A,B}$ to
$$
f_{j_A}(\omega) = \phi_-(-(jd+\delta);\omega) -
\phi_+(jd+\delta;\omega), \quad f_{j_B}(\omega) =
\phi_-(-(jd+\delta);\omega) + \phi_+(jd+\delta;\omega)
$$
and
$$
g_{j_A}(\delta) = \phi_-'(-(jd+\delta);\omega_A) -
 \phi_+'(jd+\delta;\omega_A), \quad g_{j_B}(\delta) =
\phi_-'(-(jd+\delta);\omega_B) +  \phi_+'(jd+\delta;\omega_B)
$$
for $j\in\{1,2,\ldots\}$ with $V$ still defined as in
(\ref{potentials})(ii). Non-trivial points $(\omega_*,\delta_*)$
may exist for any such $j$. For illustration we have found one
such point for $j=1$. The computed value is $(\omega_*,\delta_*)
\approx (0.73,7.33)$ and the resulting SGS corresponds to the
point $Z$ in Fig. \ref{F:SGS_homotopy_and_pt_spec_gp2}(a). Such
additional solutions are SGSs of smaller amplitude compared to
those for $j = 0$.

\begin{table}[h!]
\begin{center}
\begin{tabular}{||c||c|c|c|c|c|c||}
\hline
point & $O$ & $P$ & $Q$ & $R$ & $S$ & $T$\\
\hline
branch of solution & A & B & A & B & A & B\\
\hline
$\omega_*$ & $0.58$ & $0.70$ & $0.94$ & $0.97$ & $1.03$ &  $1.03$\\
\hline
$\delta_*$ &$1.54$ & $9.66$ & $0.78$& $3.24$ & $7.97$ & $9.57$\\
\hline
\end{tabular}
\vspace{0.25cm} \caption{Bifurcation points for surface gap
solitons in the domain $\omega \in (\omega_1,\omega_2) \cup
(\omega_3,\omega_4)$ and $\delta \in (0,d)$ for branches of solutions
given by \eqref{function-soliton}.}\label{T:bifurcations}
\end{center}
\end{table}

\subsection{Numerical Homotopy Continuation of SGSs}

Assuming the existence of a point $(\omega_*,\delta_*)$, we have
constructed the SGS of the ODE (\ref{ODE}), where
the potential function $V(x)$ is given by (\ref{potentials})(ii)
and $(\omega,\delta) = (\omega_*,\delta_*)$. The surface gap
soliton denoted as $\phi_*(x)$ is represented by one of the functions in
(\ref{function-soliton}) with $(\omega,\delta) =
(\omega_*,\delta_*)$. Each of these solutions can be used as a
starting point for a numerical homotopy continuation to generate a family
of SGS's parameterized by $\omega \subset
(\omega_{2m-1},\omega_{2m})$ for a given value of $\delta =
\delta_*$. Similarly, for a fixed $\omega=\omega_*$
a family parameterized by $\delta\subset (0,d)$ can be constructed. Under
the same assumption that the operator $\mathcal{L} = -
\partial_x^2 - \omega_* + V(x) - 3 \Gamma(x) \phi_*^2(x)$ is
invertible, the Implicit Function Theorem implies that there
exists a unique smooth continuation of the particular solution
$\phi_*(x)$ to the family of solutions along parameters $\omega$
and $\delta$.

\begin{figure}
\epsfig{figure =
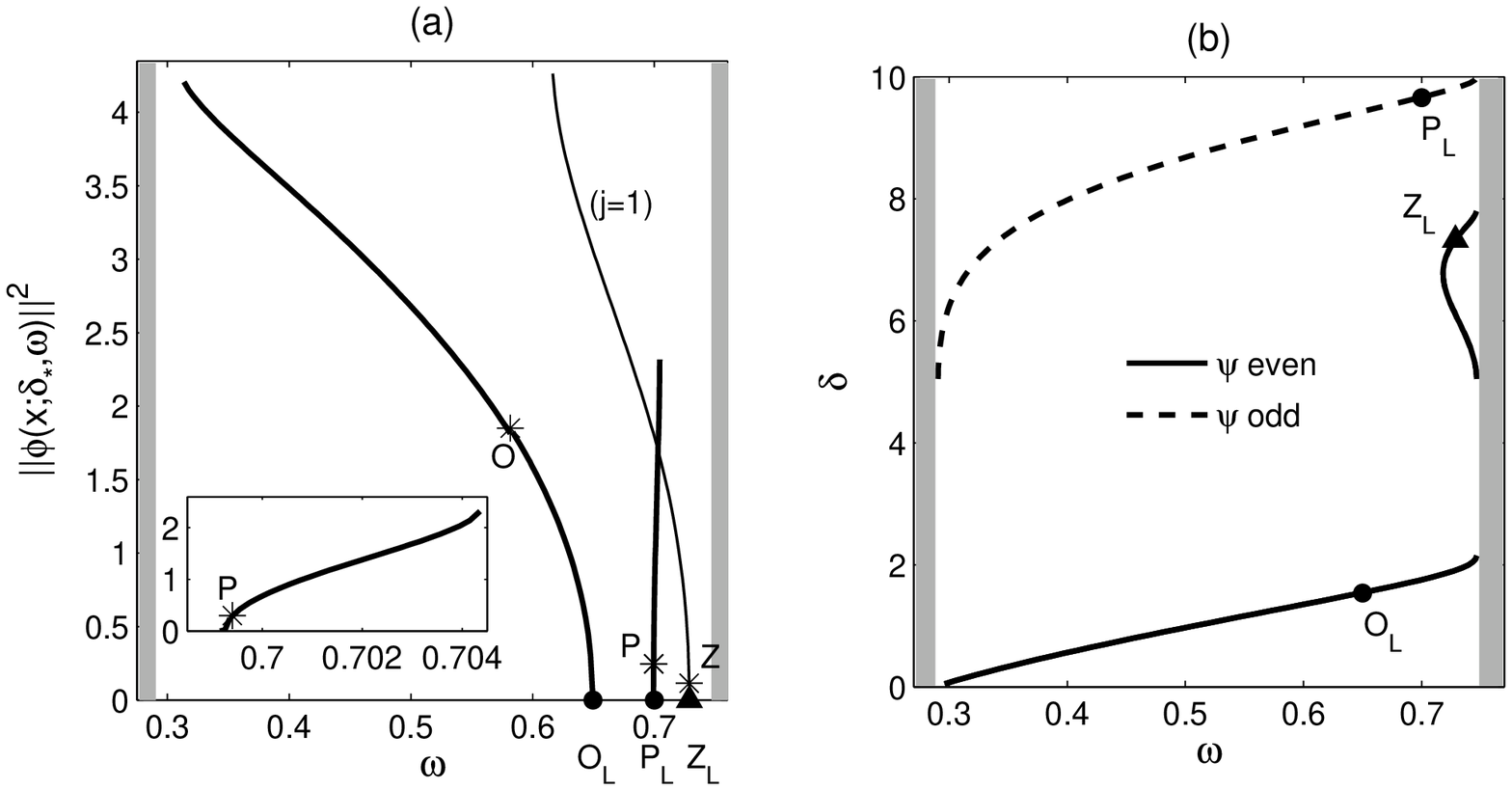,scale=.48}
\caption{(a) SGS continuation curves, total power versus
frequency, in the gap $(\omega_1,\omega_2)$. Labeled points $O,P$
correspond to those in Fig. \ref{F:om_delta_gap2} (b). Point $Z$
is discussed in Sec. \ref{S:gluing}. Points $O_L, P_L$ and $Z_L$
are SGS termination points. (b) Point spectrum of the linear
Schr\"{o}dinger operator inside  $(\omega_1,\omega_2)$ for all
$\delta \in (0,d).$ Full/dashed lines: eigenvalues with even/odd
eigenfunctions.} \label{F:SGS_homotopy_and_pt_spec_gp2}
\end{figure}

\begin{figure}
\epsfig{figure =
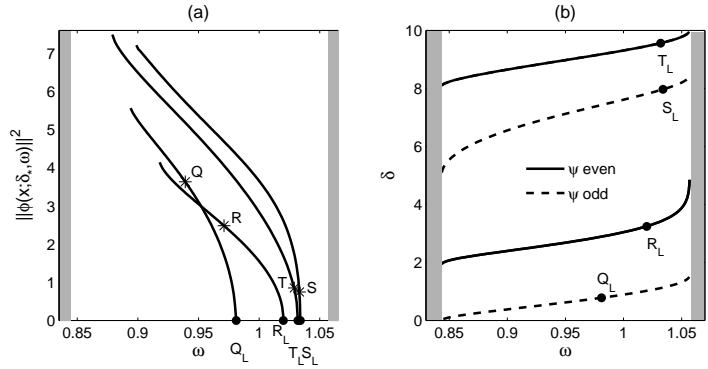,scale=.48}
\caption{(a) SGS continuation curves, total power versus
frequency, in the gap $(\omega_3,\omega_4)$. Labeled points $Q-T$
correspond to those in Fig. \ref{F:om_delta_gap3} (b). Points
$Q_L-T_L$ are SGS termination points. (b) Point spectrum of the
linear Schr\"{o}dinger operator inside  $(\omega_3,\omega_4)$ for
all $\delta \in (0,d).$ Full/dashed lines: eigenvalues with
even/odd eigenfunctions. } \label{F:SGS_homotopy_and_pt_spec_gp3}
\end{figure}

We restrict our numerical studies to the continuation in $\omega$.
Numerical results of such continuation from the SGS's at points
$O-T$ are shown in Figs. \ref{F:SGS_homotopy_and_pt_spec_gp2} (a)
and \ref{F:SGS_homotopy_and_pt_spec_gp3} (a). The curves plot the
total soliton power $\|\phi \|^2_{L^2(\mathbb{R})}$ as a function
of frequency $\omega$ for fixed $\delta = \delta_*$. Note that
each curve corresponds to a different value of $\delta_*$ and
hence a different potential $V(x)$. The values of $\delta_*$ can
be read in Table \ref{T:bifurcations}. Termination of a
continuation curve is defined when the total power of the soliton
becomes zero or when Newton iteration convergence fails. As the
figures show, the latter case is always accompanied by the slope
of the continuation curve becoming infinite suggesting a violation
of the implicit function theorem assumptions. The former
termination case is studied in the following subsection.

\subsection{Analysis of Termination Points of Surface Gap Solitons}

We shall now consider the termination points of the solution
families plotted in Figs. \ref{F:SGS_homotopy_and_pt_spec_gp2} (a)
and  \ref{F:SGS_homotopy_and_pt_spec_gp3} (a) where the soliton
power becomes zero. The points are labeled $O_L - T_L$ and their
corresponding $\delta$ and $\omega$ are given in Table
\ref{T:terminations}.
\begin{table}[h!]
\begin{center}
\begin{tabular}{||c||c|c|c|c|c|c||}
\hline
point & $O_L$ & $P_L$ & $Q_L$ & $R_L$ & $S_L$ & $T_L$\\
\hline
$\delta$ & $1.54$ & $9.66$ & $0.78$ & $3.24$ & $7.97$ & $9.57$\\
\hline
$\omega$ & $0.65$ & $0.70$ & $0.98$ & $1.02$ & $1.03$ &  $1.03$\\
\hline
\end{tabular}
\vspace{0.25cm} \caption{Termination points for the six SGS
families in  Figs. \ref{F:SGS_homotopy_and_pt_spec_gp2} (a) and
\ref{F:SGS_homotopy_and_pt_spec_gp3} (a).}\label{T:terminations}
\end{center}
\end{table}

The termination points are expected to be related to existence of non-trivial
bound states in the (point) spectrum of the Schr\"{o}dinger operator for
the same potential $V(x)$, i.e. with exponentially decaying
solutions of the linear ODE
\begin{equation}
\label{self-adjoint} - \psi'' - \omega \psi + V(x) \psi = 0,
\qquad \psi : \mathbb{R} \mapsto \mathbb{R},
\end{equation}
for $V(x)$ in (\ref{potentials})(ii) and $\omega \in
\mathbb{R}\backslash \Sigma$. The point spectrum is nonempty due to the
singularity of $V(x)$ at $x=0$.

\subsubsection{Numerical Results}

Results of numerical computations of the point spectrum contained
in the first two finite gaps $(\omega_1,\omega_2)$ and
$(\omega_3,\omega_4)$ are shown in Figs.
\ref{F:SGS_homotopy_and_pt_spec_gp2} (b) and
\ref{F:SGS_homotopy_and_pt_spec_gp3} (b) for all values $\delta
\in [0,d]$. The eigenfunctions $\psi$ are either even (full lines)
or odd (dashed lines). For the six values of $\delta$ corresponding
to the SGS families in  Figs. \ref{F:SGS_homotopy_and_pt_spec_gp2}
(a) and  \ref{F:SGS_homotopy_and_pt_spec_gp3} (a) the eigenvalues
are marked by black dots and are in perfect agreement with the
values of $\omega$ at the termination points $O_L - T_L$. The
symmetry (even/odd) of the bound states at  $O_L - T_L$ also
matches that of the eigenfunctions at the marked points in the
point spectrum. The eigenvalue curve originating as well as ending
at $\omega_2$ in  Fig. \ref{F:SGS_homotopy_and_pt_spec_gp2} (b)
corresponds to the termination point $Z_L$ of the SGS family for
$j=1$ in Fig.  \ref{F:SGS_homotopy_and_pt_spec_gp2} (a). The termination point
$Z_L$ for the same value of $\delta$ is shown by a triangle.

\subsubsection{Bifurcation Analysis for $|\delta|$ Small}

In the remaining part we consider bifurcations of point spectrum
of the Schr\"{o}dinger operator from the band edges for small
values of $|\delta|$ (or, due to the $d$ periodicity of $V$, equivalently for $\delta$ near $0$ from
above and near $d$ from below). This analysis will prove the existence of the spectral curves near
$\delta=0$ and $\delta=10$ in Figs. \ref{F:SGS_homotopy_and_pt_spec_gp2}
(b) and  \ref{F:SGS_homotopy_and_pt_spec_gp3} (b), i.e. the existence of curves with
points $O_L$ and $Q_L$ locally to $\delta=0$ and the curves with points $P_L$ and $T_L$ locally to
$\delta=10$.

In
order to construct solutions of the spectral problem
(\ref{self-adjoint}), we first consider exponentially decaying
solutions of the ODE on the half-line
$$
- \psi_+'' - \omega \psi_+ + V_0(x+\delta) \psi_+ = 0, \qquad
\psi_+ : \mathbb{R}_+ \mapsto \mathbb{R}.
$$
By using the fundamental solution of the Hill's equation
(\ref{E:spectral_prob}), we can express $\psi_+(x)$ in the form
$\psi_+ = e^{-\kappa x} u_-(x+\delta)$, where $u_-(x)$ are
periodic or anti-periodic bounded solutions of the Hill's equation
(\ref{E:spectral_prob}) with $V(x)=V_0(x)$, i.e. solutions of (\ref{E:spectral_prob})
at a band edge $\omega = \omega_n$.

As $V(x)$ is even, the function $\psi_+(x)$ admits a symmetric (even) reflection
about $x = 0$ if $\psi_+'(0) = 0$, which is equivalent to the
condition
$$
G_1(\delta,\kappa) = u_-'(\delta) - \kappa u_-(\delta) = 0,
$$
and it admits an anti-symmetric (odd) reflection
about $x = 0$ if $\psi_+(0) = 0$, which is equivalent to the
condition
$$
G_2(\delta,\kappa) = u_-(\delta) = 0.
$$
Since eigenvalues of the spectral problem (\ref{self-adjoint}) are
simple and the eigenfunctions are either even or odd, all
eigenvalues of the spectral problem (\ref{self-adjoint}) in the
band gaps $\omega \in \mathbb{R} \backslash \Sigma$ are defined by
zeros of the functions $G_1(\delta,\kappa)$ and
$G_2(\delta,\kappa)$ in $\kappa$ for a given value of $\delta$,
where $\kappa \geq 0$ and the values of $\kappa$ are related to
the values of $\omega$ in the band gaps. Both functions $G_{1,2}$ are
analytic in $\delta \in \mathbb{R}$ and periodic with period $d$.
Both functions admit analytic continuation in the parameter
$\kappa \in \mathbb{R}_+$ \cite{Kohn}.

If $\delta = 0$, the only zeros of $G_1(\delta,\kappa)$ and
$G_2(\delta,\kappa)$ occur at $\kappa = 0$, i.e. at the band edges
$\omega = \omega_n$. Indeed, if $G_1(0,\kappa) = 0$, then
$\psi_+'(0) = 0$, such that $\psi_+(x)=\psi_n(x)$ is an even function on $x \in
\mathbb{R}$. However, $\psi_+(x)$ decays exponentially as $x \to
\infty$ and grows exponentially as $x \to -\infty$ if $\kappa >
0$. Therefore, $G_1(0,\kappa) = 0$ is equivalent to $\kappa = 0$.
A similar argument works for $G_2(0,\kappa) = 0$.

\medskip
\noindent
\textit{(i) Bifurcation of Even Eigenfunctions}

Let us first consider the zeros of $G_1(\delta,\kappa)$. Computing the derivatives of $G_1(\delta,\kappa)$ in $\delta$ and
$\kappa$ at $(\delta,\kappa) = (0,0)$, we obtain
\begin{eqnarray*}
\partial_{\delta} G_1(0,0) & = & u_-''(0) = \psi_n''(0) = (V_0(0) -
\omega_n) \psi_n(0), \\ \partial_{\kappa} G_1(0,0) & = &
-\tilde{\psi}_n'(0) - \psi_n(0),
\end{eqnarray*}
where $\tilde{\psi}_n$ is the generalized Bloch function, see \eqref{E:Bloch_fn_eqs}. The fact
$\tilde{\psi}_n = -\left.\frac{\partial u_-}{\partial \kappa}\right|_{\kappa=0}$ is clear from
differentiation of \eqref{E:spectral_prob} with respect to $\kappa$.

It is found in \cite{Pelin} that
$$
D(x) = \psi_n(x) \tilde{\psi}_n'(x) - \psi_n'(x) \tilde{\psi}_n(x)
+ \psi_n^2(x)
$$
is constant in $x$, i.e. $D(x) =D(0)$, and that
\begin{equation}\label{E:D_of_zero}
D(0) = \frac{1}{2} \omega_{2n-1,2n}''(k) (\psi_n,\psi_n),
\end{equation}
where either $k = 0$ or $k = \frac{\pi}{d}$ at the bifurcation
point $\omega = \omega_n$. Since $\psi_n'(0)=0$,
$D(0)=\psi_n(0)(\tilde{\psi}'_n(0)+\psi_n(0))$ and the
leading-order approximation for the root of $G_1(\delta,\kappa)$
near $(\delta,\kappa) = (0,0)$ is given by
$$
\delta = \frac{\tilde{\psi}_n'(0) + \psi_n(0)}{\psi_n(0)(V_0(0) - \omega_n)
} \kappa + \Order(\kappa^2) = \frac{D(0)}{
\psi_n^2(0) (V_0(0)-\omega_n)} \kappa + \Order(\kappa^2),
$$
where $\psi_n(0) \neq 0$ (since $\psi_n'(0) = 0$).  Using \eqref{E:D_of_zero}
and the facts $\omega_n > 0$ and $V_0(0)=0$ for the numerical example
(\ref{potential-example}), we get
$$
\delta = -\frac{\omega_{2n-1,2n}''(k) (\psi_n,\psi_n)}{2
\psi_n^2(0) \omega_n} \kappa + \Order(\kappa^2).
$$
Therefore, the bifurcation occurs for $\delta > 0$ if
$\omega_{2n-1,2n}''(k) < 0$ (e.g. for $\omega$ to the right of
$\omega_1$) and for $\delta < 0$ if $\omega_{2n-1,2n}''(k) > 0$
(e.g. for $\omega$ to the left of $\omega_0$ and
$\omega_4$), see Table 1. Note that the negative values of
$\delta$ correspond to the values of $\delta$ below the level
$\delta = d$ due to periodicity of the function
$G_1(\delta,\kappa)$ in $\delta$. The above local existence
analysis for even bound states is confirmed by the full lines near
$\delta=0$ in Fig. \ref{F:SGS_homotopy_and_pt_spec_gp2} (b) and
near $\delta=d=10$ in Fig. \ref{F:SGS_homotopy_and_pt_spec_gp3}
(b).

\medskip
\noindent
\textit{(ii) Bifurcation of Odd Eigenfunctions}

 Similarly to (i), we study the zeros of $G_2(\delta,\kappa)$. We
compute the derivatives of $G_2(\delta,\kappa)$ in $\delta$ and
$\kappa$ at $(\delta,\kappa) = (0,0)$
\begin{eqnarray*}
\partial_{\delta} G_2(0,0) & = & u_-'(0) = \psi_n'(0), \\
\partial_{\kappa} G_2(0,0) & = & -\tilde{\psi}_n(0),
\end{eqnarray*}
such that the leading-order approximation for the root of
$G_2(\delta,\kappa)$ near $(\delta,\kappa) = (0,0)$ is given by
$$
\delta = \frac{\tilde{\psi}_n(0)}{\psi_n'(0)} \kappa +
\Order(\kappa^2) = -\frac{\omega_{2n-1,2n}''(k) (\psi_n,\psi_n)}{2
(\psi_n'(0))^2} \kappa + \Order(\kappa^2),
$$
where $\psi_n'(0) \neq 0$ and $D(0)=-\psi_n'(0)\tilde{\psi}_n(0)$
(since $\psi_n(0) = 0$). Therefore, the bifurcation occurs for
$\delta > 0$ if $\omega_{2n-1,2n}''(k) < 0$ (e.g. for $\omega$ to
the right of $\omega_3$) and for $\delta < 0$ if
$\omega_{2n-1,2n}''(k) > 0$ (e.g. for $\omega$ to the left of
$\omega_2$). The dashed lines near $\delta=0$ in Fig.
\ref{F:SGS_homotopy_and_pt_spec_gp3} (b) and near $\delta=d=10$ in
Fig. \ref{F:SGS_homotopy_and_pt_spec_gp2} (b) confirm this
analysis.

Note that there are curves in Figs.
\ref{F:SGS_homotopy_and_pt_spec_gp2} (b) and
\ref{F:SGS_homotopy_and_pt_spec_gp3} (b) which do not bifurcate
from $\delta = 0$ and $\delta = d = 10$ but still bifurcate from
the band edge $\omega = \omega_n$. Bifurcations of these curves
cannot be confirmed from the analytical theory above, unless the
values of $G_{1,2}(\delta;0)$ for $0 < \delta < d$ are
approximated numerically.

\section{Conclusion}
\label{S:surface-conclusion}

We have employed methods of bifurcation theory for the existence
problem of SGS's supported by the nonlinearity interface and the
periodic potential. Two bifurcation problems are considered
numerically. The first bifurcation takes place from the standard
gap solitons existing at the zero jump of the nonlinearity
coefficient. The second bifurcation takes place from the bound
state consisting of parts of two standard gap solitons glued
together in a continuously differentiable SGS. Three asymptotic
results are described in the article. We show that the standard
gap solitons can be continued generally for small jumps in the
nonlinearity coefficient. On the contrary, no SGS's for non-zero
jump of the nonlinearity coefficient exists in the NLS
approximation which is valid near the band edges. In addition, we
study analytically bifurcations of eigenvalues of the
Schr\"{o}dinger operator with non-smooth potential from band edges
of the Hill's equation.

One can argue that the SGS's bifurcating from a
standard gap soliton or a gluing combination of two gap solitons
inherits stability properties of gap solitons in the neighborhood
of the local bifurcation points. Stability of standard gap
solitons was considered analytically and numerically in
\cite{PSK04}. The stability properties can change far from the
bifurcation points. Detailed computatitons of stability of the
SGS's will be the subject of the forthcoming work.

{\bf Acknowledgement.} T.D. is supported by ETH Research
Fellowship. D.P. is supported by the Humboldt Research Fellowship
hosted at Institut f\"{u}r Analysis, Dynamik und Modellierung,
Fakultaet f\"{u}r Mathematik und Physik at the Universit\"{a}t
Stuttgart. He thanks people at ETH Zurich for hospitality during
his visit.


\begin{thebibliography}{99}

\bibitem{Blomer} D. Bl\"{o}mer, A. Szameit, F. Dreisow, T. Schreiber, S. Nolte, and A. T\"{u}nnermann,
``Nonlinear refractive index of fs-laser-written waveguides in
fused silica," Opt. Express {\bf 14}, 2151-2157 (2006)

\bibitem{Eastham73} M.S. Eastham, {\em The Spectral Theory of Periodic Differential Equations},
(Scottish Academic Press, Edinburgh, 1973)

\bibitem{FSW06} G. Fibich, Y. Sivan, and  M. I. Weinstein, ``Bound states of nonlinear Schr\"{o}dinger equations with
a periodic nonlinear microstructure,'' Physica D {\bf 217}, 31--57
(2006)

\bibitem{HSCS05} J. Hudock, S. Suntsov, D. Christodoulides, and G. Stegeman,
``Vector discrete nonlinear surface waves," Opt. Express {\bf 13},
7720--7725 (2005)

\bibitem{KEVT06} Y.V. Kartashov, A.A. Egorov, V.A. Vysloukh, and L. Torner, ``Surface vortex solitons," Opt. Express
{\bf 14}, 4049--4057 (2006)

\bibitem{KT06} Y.V. Kartashov and L. Torner, ``Multipole-mode surface solitons," Opt. Lett. {\bf 31}, 2172--2174 (2006).

\bibitem{KVT06} Y.V. Kartashov, V.A. Vysloukh and L. Torner, ``Surface gap solitons",
Phys. Rev. Lett. {\bf 96}, 073901 (2006)

\bibitem{Kohn} W. Kohn, "Analytic properties of Bloch waves and
Wannier functions", Phys. Rev. {\bf 115}, 809--821 (1959)

\bibitem{Magnus_Win_66} W. Magnus and S. Winkler, {\em Hill's equation}, Interscience Tracts in
Pure and Applied Mathematics, No. {\bf 20} (John Wiley \& Sons,
New York-London-Sydney, 1966)

\bibitem{MSCSH05} K.G. Makris, S. Suntsov, D.N. Christodoulides, G.I. Stegeman, and A. Hache,
``Discrete surface solitons," Opt. Lett. {\bf 30}, 2466--2468
(2005)

\bibitem{pankov} A. Pankov, "Periodic nonlinear Schrödinger equation with
application to photonic crystals", Milan J. Math. {\bf 73},
259--287 (2005)

\bibitem{Pelin} D. Pelinovsky, "Asymptotic reductions of the Gross--Pitaevskii
equation", in {\em Emergent nonlinear phenomena in Bose--Einstein
condensates: Theory and Experiment}, Eds. P. Kevrekidis, D.
Frantzeskakis, and R. Carretero (Springer, Heidelberg, 2007)

\bibitem{PSK04} D.E. Pelinovsky, A.A. Sukhorukov, and Y. Kivshar,
``Bifurcations and stability of gap solitons in periodic
structures," Phys. Rev. E {\bf 70}, 036618 (2004)

\bibitem{Kato} B. Simon and M. Reed, {\em Methods of Modern
Mathematical Physics IV: Analysis of Operators}, (Academic Press,
New York, 1978)

\bibitem{FSW07} Y. Sivan, G. Fibich, and M. I. Weinstein, ``Waves in nonlinear lattices -
ultrashort optical pulses and Bose--Einstein condensates,'' Phys.
Rev. Lett. {\bf 97}, 193902 (2006)

\bibitem{SMCSHMYSS06} S. Suntsov, K. G. Makris, D. N. Christodoulides, G. I. Stegeman,
A. Hache, R. Morandotti, H. Yang, G. Salamo, and M. Sorel,
``Observation of Discrete Surface Solitons,'' Phys. Rev. Lett.
{\bf 96}, 063901 (2006)

\bibitem{T80} W. J. Tomlinson, ``Surface wave at a nonlinear interface," Opt. Lett. {\bf 5}, 323--325 (1980)

\end{thebibliography}
\end{document}